\documentclass[12pt]{article}

\usepackage[pdftex]{graphicx}

\usepackage{titlesec}
\usepackage[titletoc,title]{appendix}
\usepackage{caption}
\usepackage{subcaption}
\usepackage{dcolumn}
\usepackage{bm}
\usepackage{amsfonts}
\usepackage{amsmath}
\usepackage{amssymb}
\usepackage{hyperref}

\usepackage{adjustbox}
\usepackage{array}

\usepackage{color}
\usepackage{textcomp}

\usepackage[letterpaper, total={7in, 8in}]{geometry}

\def\be{\begin{equation}}
\def\ee{\end{equation}}

\newcommand{\rmd}{\mathrm{d}}

\newcommand{\avg}[1]{\langle #1 \rangle}

\renewcommand{\[}{\left[}
\renewcommand{\]}{\right]}
\renewcommand{\(}{\left(}
\renewcommand{\)}{\right)}

\newcommand{\sig}{\sigma}

\newcommand{\bea}{\begin{eqnarray}}
\newcommand{\eea}{\end{eqnarray}}

\newcommand{\bi}{\begin{itemize}}
\newcommand{\ei}{\end{itemize}}

\newcommand{\beq}{\begin{equation}}
\newcommand{\eeq}{\end{equation}}
\newcommand{\beqa}{\begin{eqnarray}}
\newcommand{\eeqa}{\end{eqnarray}}

\begin{document}

\begin{titlepage}

\begin{flushright}
    CERN-TH-2016-228
\end{flushright}

\bigskip\

\vspace{.5cm}
\begin{center}

{\fontsize{25}{28}\selectfont  \sffamily \bfseries  A Theory of Taxonomy}

\end{center}

\vspace{1cm}

\begin{center}
{\fontsize{12}{30}\selectfont Guido D'Amico$^{\diamondsuit \spadesuit}$\footnote{damico.guido@gmail.com},
Raul Rabadan$^\heartsuit$\footnote{rr2579@cumc.columbia.edu}, and Matthew Kleban$^\spadesuit$\footnote{kleban@nyu.edu}} \end{center}

\begin{center}

\vskip 8pt

\textsl{$^\diamondsuit$ Theoretical Physics Department, CERN, Geneva, Switzerland}
\vskip 7pt

\textsl{$^\spadesuit$ Center for Cosmology and Particle Physics, Physics Department, New York University, New York, USA}
\vskip 7pt

\textsl{$^\heartsuit$ Departments of Systems Biology and Biomedical Informatics, Columbia University, New York, USA}
\vskip 7pt

\end{center}

\vspace{1.2cm}
\hrule \vspace{0.3cm}
\noindent {\sffamily \bfseries Abstract} \\[0.1cm]
A taxonomy is a standardized framework to classify and organize
items into categories.
Hierarchical taxonomies are ubiquitous, ranging from the classification of organisms to the file system on a computer.
Characterizing the typical distribution of items within taxonomic categories is an important question with applications in many disciplines.
Ecologists have long sought to account for the patterns observed in species-abundance distributions (the number of individuals per species found in some sample)~\cite{Alroy:2015hh,Bowler:2012ce}, and computer scientists study the distribution of files per directory~\cite{Agrawal:2007if}.
Is there a universal statistical distribution describing how many items are typically found in each category in large taxonomies?
Here, we analyze a wide array of large, real-world datasets -- including items lost and found on the New York City transit system, library books, and a bacterial microbiome -- and discover such an underlying commonality.
A simple, non-parametric branching model that randomly categorizes items and takes as input only the total number of items and the total number of categories successfully reproduces the abundance distributions in these datasets.
This result may shed light on patterns in species-abundance distributions long observed in ecology~\cite{Preston:1948he}.
The model also predicts the number of taxonomic categories that remain unrepresented in a finite sample.
\vskip 10pt
\hrule

\vspace{0.6cm}
\end{titlepage}

\begin{figure}
    \centering
    \begin{subfigure}[t]{0.35\textwidth}
        \includegraphics[width=\textwidth]{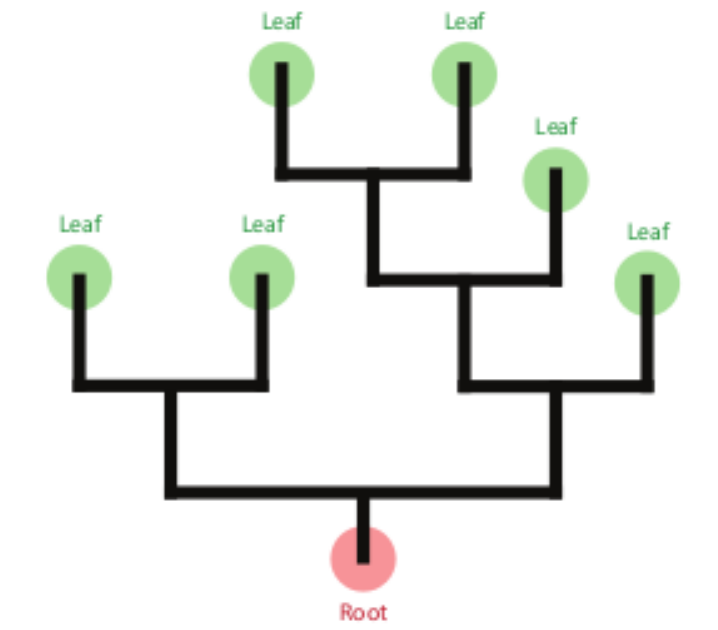}
        \caption{Representation of a binary hierarchical tree.}
        \label{fig:tree}
    \end{subfigure}
    \begin{subfigure}[t]{0.55\textwidth}
        \includegraphics[width=\textwidth]{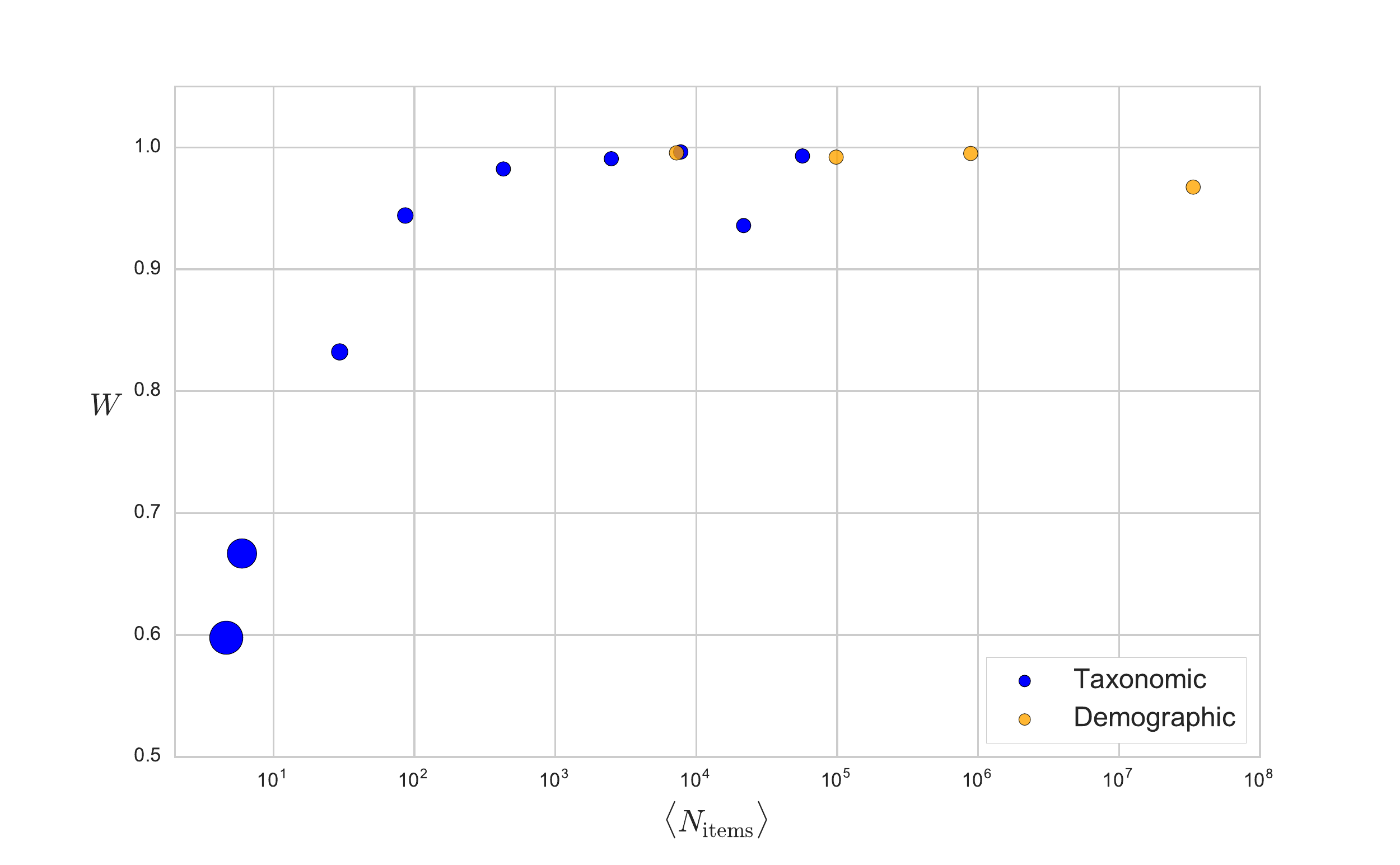}
        \caption{Degree of (log)normality of each dataset as measured by the Shapiro-Wilk statistic $W$ versus $n/k$, the average number of items per category.
        The size of the dot scales with $q/k$, a measure of how poorly-sampled the dataset is that is strongly correlated with $W$ (Pearson correlation between $\log(q/k)$ and $log(W)$ is $-0.9532$ with p-value $1.98\times 10^{-5}$;
        Spearman correlation between $\log(q/k)$ and $log(W)$ is $-0.5829$ with p-value $0.077$).}
        \label{fig:plotW}
    \end{subfigure}
    \begin{subfigure}[t]{0.8\textwidth}
        \includegraphics[width=\textwidth]{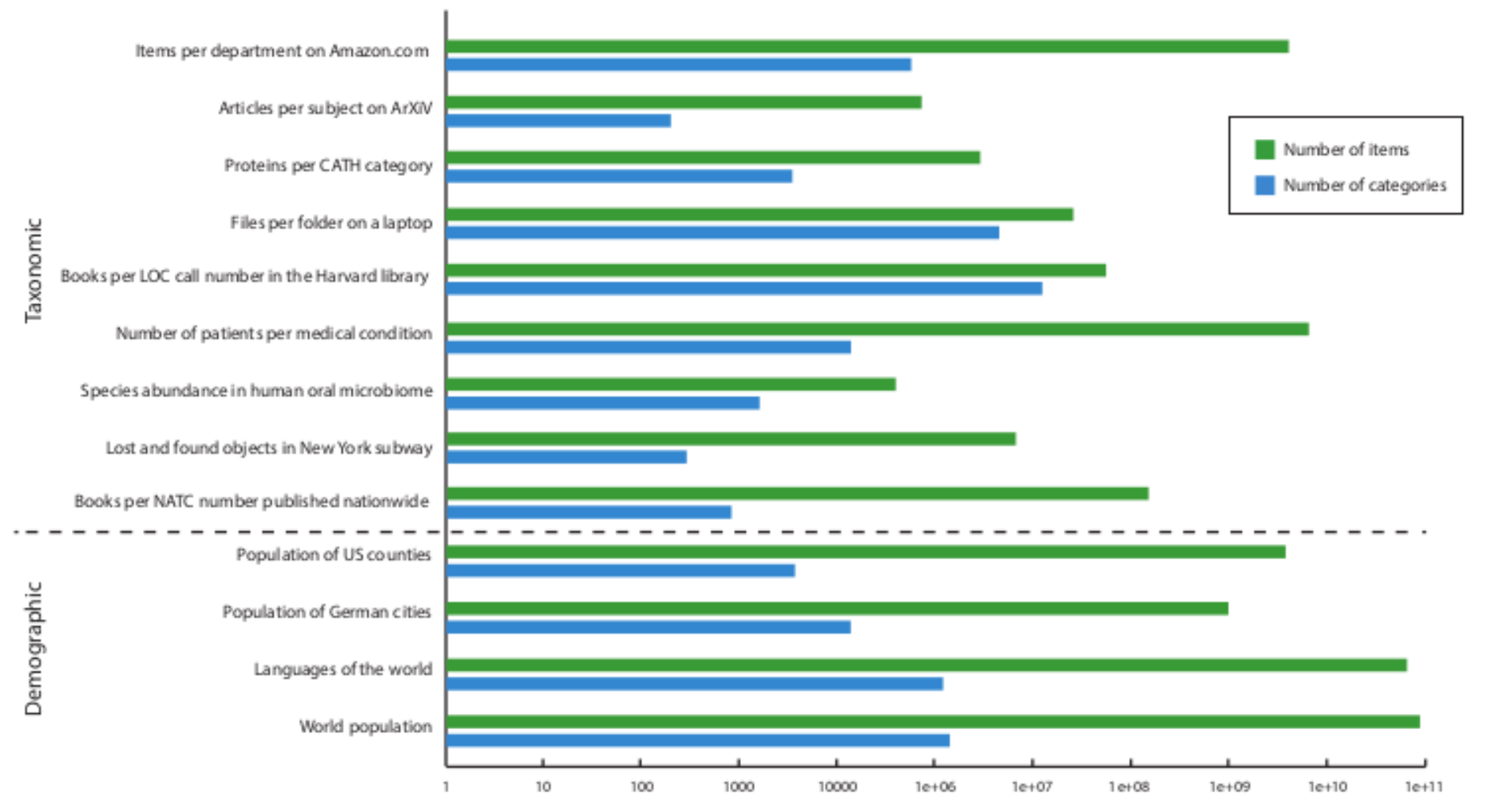}
        \caption{The datasets used in this analysis.
        Blue bars indicate the number of categories $k$, green the total number of categorized items $n$.}
        \label{fig:alldata}
    \end{subfigure}
    \caption{Hierarchical taxonomies and the datasets used in this analysis.}
\end{figure}

Hierarchical taxonomies are tree-like structures that organize items into subcategories, each of which descends along a unique path from a root category containing all items (Fig.~\ref{fig:tree}).
From Aristotle to the Linnean classification of species, taxonomies have long been part of our way of classifying organisms, and more generally large corpora of knowledge.

We study the distribution of items among categories in a variety of large hierarchical classification systems, including disease incidence in 250 million patients~\cite{Hripcsak:2016fu}, bacterial microbiomes~\cite{Dewhirst:2010iv}, items for sale on \href{www.amazon.com}{www.amazon.com}, books in the Harvard University library system, files per directory on a laptop, and items lost and found on the New York City transit system (Fig.~\ref{fig:alldata}, Table~\ref{tab:data}, and Appendix \ref{sec:data_description}).
In all these datasets, a few categories are very popular while many categories contain only a few items.
When categories are sorted by number of items and binned into a histogram, one can observe a range of shapes (Fig.~\ref{fig:plotW}).
As a function of the logarithm of the number of items, many datasets present a characteristic Gaussian or bell-shaped form -- for patients per medical condition and \href{www.amazon.com}{www.amazon.com}'s categorization -- while others tend towards a highly skewed distribution with a long tail corresponding to many categories with low number of items -- microbiome, files per directory, and the Harvard library books.
This variation is correlated to how well-sampled the distribution is: datasets with a sufficiently large number of items relative to the number of categories are close to Gaussian, while those with many categories containing only a few items are skewed (Fig.~\ref{fig:plotW}).

\begin{figure}[h]
    \centering
    \includegraphics[width=0.7\textwidth]{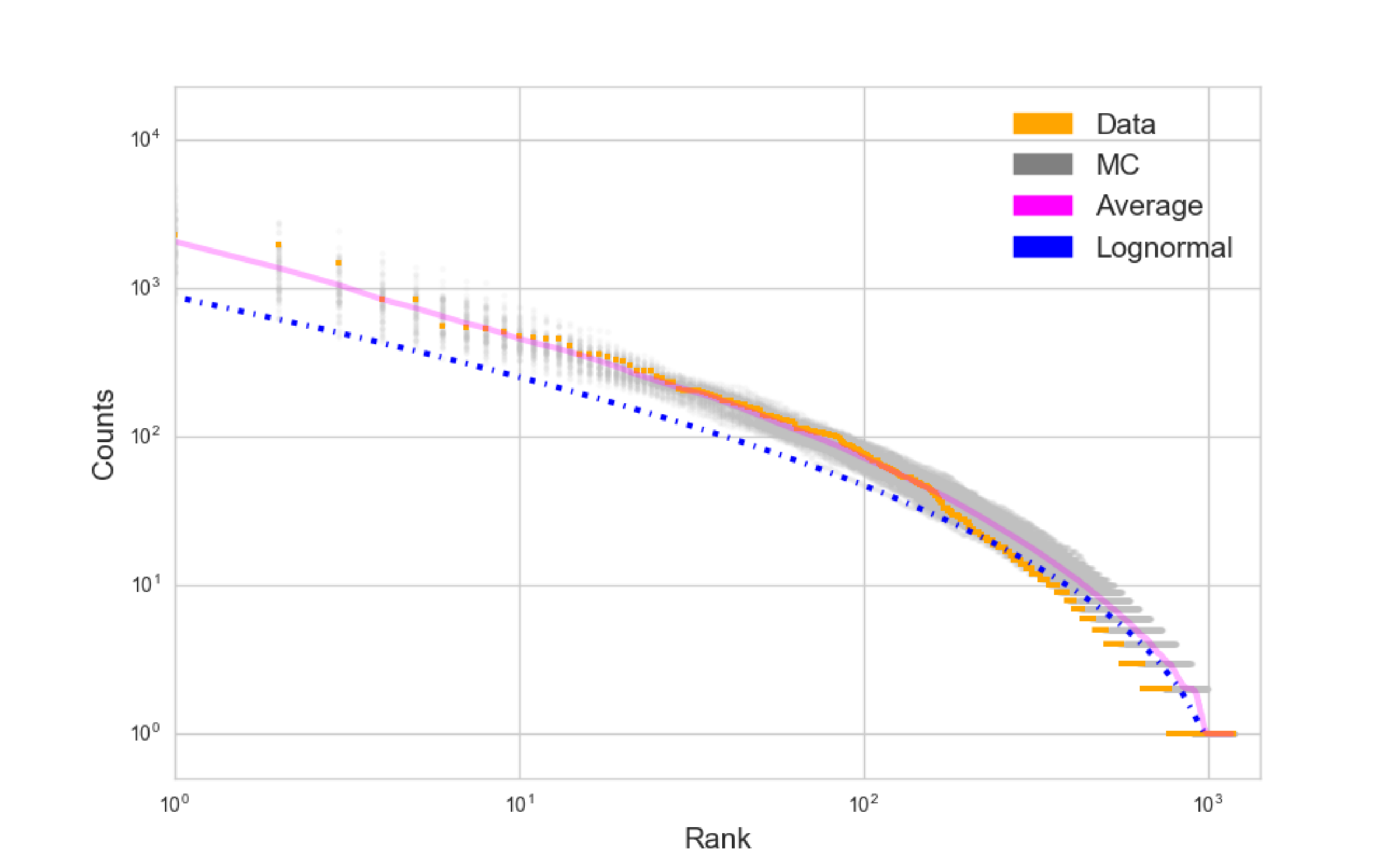}
    \caption{Rank-abundance of human oral microbiome data}
    \label{fig:microbiome}
\end{figure}

The archetypal taxonomy is the categorization of organisms into species.
Characterizing the distribution of the number of individuals of each species found in some area or sample -- the so-called species abundance distribution (SAD) -- is a long standing problem in
ecology~\cite{Fisher:1943kj}.
A classic observation is that SADs follow a ``canonical lognormal distribution'', a lognormal distribution with standard deviation $\sigma$ determined by the number of species $k$~\cite{Preston:1948he,Sugihara:1980bx}:
\be
    k = \frac{\sig}{\ln 2} \sqrt{\frac{\pi}{2}} \exp{\sig^2/2} \, .
    \label{eq:canln}
\ee
However, in many real-world examples neither the canonical lognormal nor a lognormal with mean and variance chosen to best fit the data describe real-world SADs particularly well~\cite{Alroy:2015hh,Harte:1999ir}.
In Figure~\ref{fig:microbiome} we present such a dataset: a microbiome (a sample of a bacterial population categorized using Operational Taxonomic Units) for which the best-fit lognormal is a poor fit.
\begin{figure}
    \includegraphics[width=0.95\textwidth]{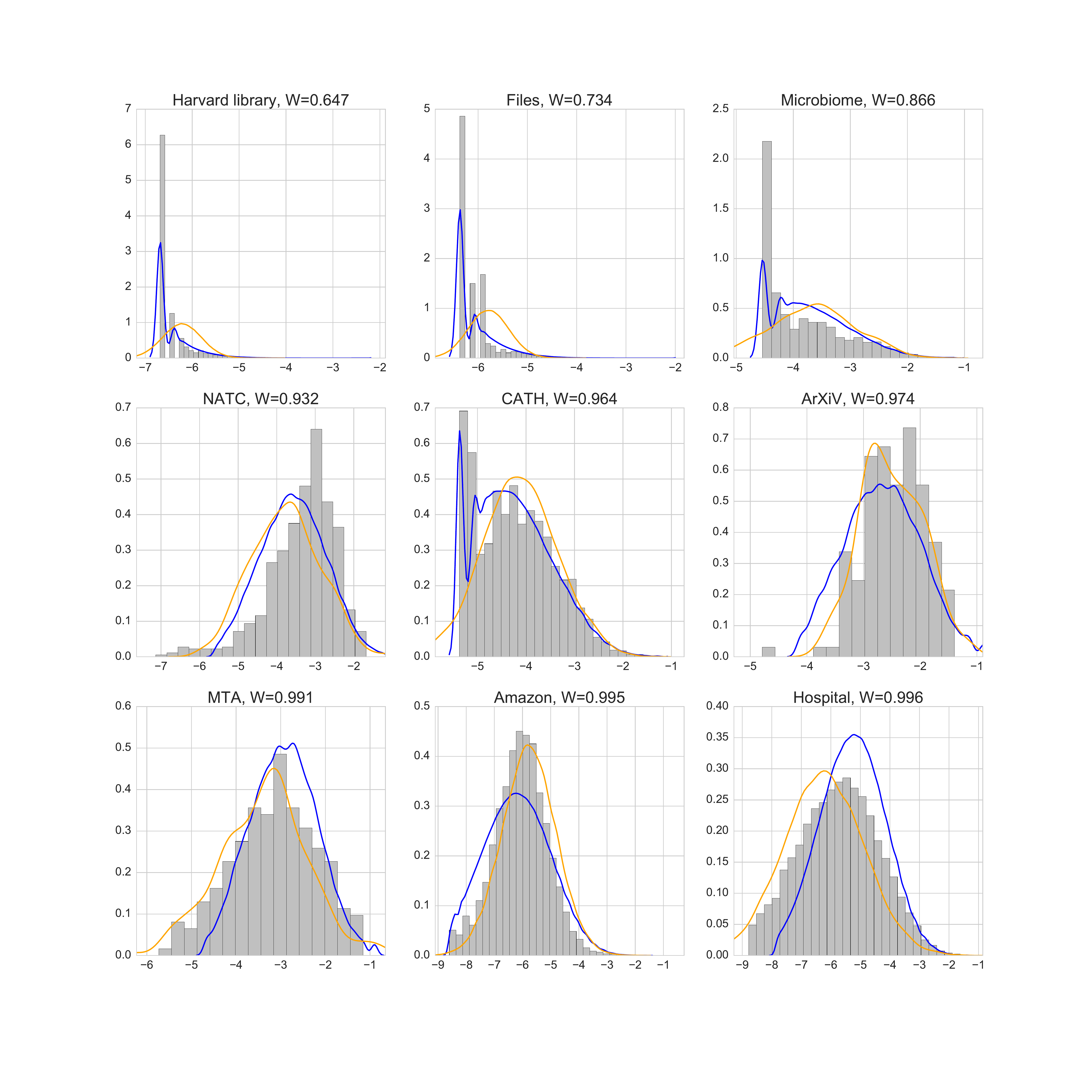}
    \caption{Histograms of the hierarchical taxonomic datasets.
    The datasets are ordered left-right and top-bottom by increasing value of the Shapiro-Wilk statistic (measuring degree of normality).
    Grey is data, and we show kernel density estimates of the average of the random categorization model (blue curves) and the best fit lognormal (orange curves).
    In each plot the horizontal axis is the log of the fraction of items in the category, and the vertical axis is the log of the number of categories in each bin.}
    \label{fig:histograms_sw}
\end{figure}
In ecology, Sugihara~\cite{Sugihara:1980bx} proposed a ``sequential broken stick'' model to explain the origin of~\eqref{eq:canln}.
The model randomly and sequentially divides items into subcategories, producing a branching tree.
Sugihara's approach to SAD was motivated by ecological niches and evolution, and he argued against a purely statistical explanation.
However, if any set of items originally in one category are repeatedly and randomly subdivided, by the central limit theorem the logarithm of the number in any given subcategory should approach a normal distribution in the limit of many divisions, with a variance that depends on the number of divisions in a way closely related to~\eqref{eq:canln}.
For this reason Sugihara's model produces an approximation to a lognormal distribution, but the same would hold for any random process involving repeated subdivisions~\cite{Drmota:2009tm}.

Our hypothesis is that sequential random categorization can successfully account for the item abundance distribution in large hierarchical taxonomies, including those that have nothing to do with ecology.
Random categorization can account for both the lognormal distributions observed in well-sampled datasets and the skewed shape of more poorly sampled taxonomies.
In addition it can predict the detailed shape of the distribution (including its mean, variance, and the number of unsampled, empty categories) based only on the number of items and categories, entirely independently of the nature of the items being categorized.

Here, we present and analyze what we regard as the simplest random categorization model (see Appendix \ref{model} for details).
It is distinguished by the fact that within a broad class of related models, it produces the item distribution with the minimum possible variance.
The only inputs are the total number of categories $k$ and total number of items $n$, and it contains no continuous or adjustable parameters.

The model divides a single root category into two subcategories, then randomly selects one subcategory and divides it into two, etc., producing a random binary tree (Fig.~\eqref{fig:tree}).
After $q-1$ divisions the tree has $q$ ``leaves'', each of which represents a category. Probabilities are then assigned to the leaves in a way that depends in a simple way on the number of branchings separating the leaf from the root.
Finally, we ``sample'' the tree by populating the leaves with $n$ items, distributed randomly according to the leaf probabilities.

The motivation for this last step is as follows.
We regard the tree as representing a kind of ``ideal'' classification, while any real dataset is a sampling of that categorization.
An example is the Library of Congress (LOC) categorization system, with $q$ total categories.
The collection of any individual library is a sampling of the full LOC collection.
In most real libraries the sampling is sparse and only $k < q$ of the LOC categories are represented by books in that library's collection.
If the number of items $n$ is sufficiently large so that there are few or no empty categories, the model predicts an item abundance distribution that is indeed approximately lognormal (for reasons related to the central limit theorem and the statistics of the tree distribution).
For smaller $n$, the distribution of items in non-empty categories differs substantially from lognormal.
In this way the same model -- one with no continuous parameters and no fitting to the data -- can produce distributions that look very different in different regimes, successfully reproducing the wide range represented in Figure~\ref{fig:histograms_sw}.

Generally, from the dataset we only know the number of non-empty categories $k$.
To model the data, we therefore generate trees with $q \geq k$ categories.
We choose $q$ so that the typical number of non-empty categories is $k$ when the tree is sampled with $n$ items (see Methods for details).
For each dataset, the raw data -- number of items per category -- is sorted into a ranked list.
The category with the largest number of items is ranked $1$, while that with the least is ranked $k$.
We then plot category rank versus the number of items (Fig.~\ref{fig:microbiome}), or a binned histogram of the number of categories with some range of numbers of items (Fig.~\ref{fig:histograms_sw}).
The former is closely related to a graph of the cumulative distribution function of the item partitioning distribution, because for a category with rank $r$, $r/k$ is the probability that a random category would have as many or more items.
The latter is related to the probability distribution function for the item distribution.

The model can be thought of as producing a random partition of $n$ items into $k$.
To compare the model to a particular dataset we use the model to generate many such partitions, and compute the relative entropy or Kullback-Leibler (KL) divergence of each to the data, generating a distribution of KL statistics (Appendix \ref{comp}).
The KL statistic is a measure of how informative the model is about the data -- if it is small, it indicates that the model contains most of the information in the data.
To compare our model to lognormal, for each dataset we also compute the KL statistics for partitions generated by $k$ random draws from a lognormal distribution with mean and variance fit to the data using maximum likelihood estimators (MLE).
In most cases -- including some of the best-sampled datasets -- according to this measure the tree model outperforms the best-fit lognormal (Fig.~\ref{fig:allKL}), which is quite remarkable given its lack of adjustable parameters.
This suggests that the decisions made in designing hierarchical categorization systems follow a general pattern that our random categorization model is capturing.

As a test, we compare the model to four categorizations that are not hierarchical taxonomies: populations of German cities, U.S. counties, world countries, and the number of speakers per language.
In contrast to the hierarchical systems, the model is clearly inferior to the best-fit lognormal in all cases except for world population, where it is marginally inferior.

An important point is that $q$ can be thought of as a \emph{prediction} for the total number of categories -- that is, $q-k$ is the number of categories that will \emph{not} be represented in a typical sample of $n$ items.
This type of prediction has utility in a broad array of circumstances, and our model's prediction exhibits a clear correlation (Fig. S3) to commonly employed estimators such as Chao~\cite{Chao:1984vh}.
An interesting test case is the Harvard library collection.
The ratio of the total number of existing Library of Congress categories to $k$ (the number of categories represented in Harvard's collection) is approximately 16.
This constitutes a lower bound on the true ratio, in the sense that the Library of Congress system may still be incomplete (certain subjects may remain as-yet unclassified, due to the dearth of extant books on those topics).
Our model predicts a ratio of total categories to non-empty categories $q/k = 38$, while the Chao estimator predicts a ratio of 2.6.
Hence, Chao dramatically under-predicts the number of empty categories.
Our model may over-predict it, or it may correctly capture the current incomplete state of the system.

Large hierarchical classification systems are ubiquitous, and our results show a distinct and novel pattern in the distribution of items among categories in a broad array of such systems.
Randomly categorizing items in a binary tree, with no adjustable parameters and no input regarding the type of items being classified, successfully reproduces this pattern.
This result has broad applicability across many fields of science, and should serve as a starting point -- a null hypothesis -- for more detailed investigations of item distributions.
Deviations from the pattern observed here likely indicate that there is some significant effect that is not captured by random categorization -- for instance preferential attachment~\cite{Albert:1999co,Albert:2002it}, where added items are substantially more (or less) likely to be assigned to categories that are already large.
The model can also be used to predict the number of empty or missing categories: given a finite sample of items, it makes a prediction for how many more categories there are than those represented in the sample.
An important application of this prediction arises in ecology, where it can be used to estimate the number of rare species that were missed in some survey, and identifying the number of such ``missing categories'' clearly has applications far beyond ecology~\cite{Lynch:1965dd}.
Another feature of interest is that many of these datasets appear very close to power law over a large range of category ranks (Fig.~\ref{fig:microbiome}), and this behavior is correctly reproduced by the model.
This may help account for the power laws previously observed in many contexts even without preferential attachment~\cite{Zipf:1935uq}.

\paragraph{Acknowledgments}
The work of MK is supported in part by the NSF through grant PHY-1214302, and he acknowledges membership at the NYU-ECNU Joint Physics Research Institute in Shanghai.
RR is supported by NIH grants U54-CA193313-01 and R01-GM117591-01.
GDA acknowledges the Aspen Center for Physics for hospitality during some of this work.
We gratefully acknowledge Elisa Araldi, Joel Cohen, Kyle Cranmer, Valentino Foit, Paul Ginsparg, David Hogg, and Chuck Newman for stimulating discussions, and Tim Chu for helping with figures.

\titleformat{\section}{\large\bfseries}{\appendixname~\thesection :}{0.5em}{}

\begin{appendices}

\section{Data description}
\label{sec:data_description}

Our analysis makes use of ten datasets.
From each we extracted a list of categories and ranked it according to the number of items in each category (category 1 has the most items, category $k$ the fewest).
Here we provide a brief description of each dataset and its source.
A summary can be found in Table~\ref{tab:data}.

\begin{enumerate}

\item{Harvard library:}
   We counted the number of books per Library of Congress call number, extracted from the Harvard Library dataset published on \url{http://library.harvard.edu/open-metadata\#Harvard-Library-Bibliographic-Dataset}.
   We considered call numbers that differed only after the ``.'' as equivalent.
\item{Files:}
  We counted the number of files in each directory on the Macbook Air laptop computer of the corresponding author, using the Bash command \url{find . -type d | while read -r dir; do find "$dir" -type f -maxdepth 1 | wc -l; done >> filedat}.
  We checked that the file systems on several other Macintosh computers have a similar structure.
\item{Microbiome:}
  We used rank abundance of bacteria per operational taxonomic unit in a sample of the human oral microbiome, as described in \url{http://jb.asm.org/content/192/19/5002.full\#sec-14}~\cite{Dewhirst:2010iv}.
\item{NATC:}
  The number of books per category in the North American Title Count database, using a file sent to us by a librarian at the Library of Congress.
  The file is available upon request.
\item{CATH:}
  We took data from the file \url{http://release.cathdb.info/latest_release/CathDomainList.v4.0.0},
  which is a database of protein structure classification.
  We define as a category the hierarchy of the first four levels of the classification, as described in \url{http://www.ncbi.nlm.nih.gov/pmc/articles/PMC3525972/}, and then count the number of domains in each category.
\item{ArXiv:}
  Using public APIs from the website \href{http://www.arxiv.org}{www.arxiv.org}, we counted the number of articles published in each subcategory in the year 2014.
  We considered only the primary category of each article, neglecting cross-listings.
\item{MTA:}
  We counted the number of items per category in the ``lost and found'' inventory of the Metropolitan Transportation Authority of New York City, as of December 24\textsuperscript{th}, 2015.
  The data are published and constantly updated on \href{http://advisory.mtanyct.info/LPUWebServices/CurrentLostProperty.aspx}{http://advisory.mtanyct.info/LPUWebServices/CurrentLostProperty.aspx}, and we used the dataset downloaded and published on \href{https://github.com/atmccann/mta-lost-found/blob/master/data/2014-12-24.json}{https://github.com/atmccann/mta-lost-found/blob/master/data/2014-12-24.json}
\item{Amazon:}
  We counted the number of items for sale on in departments with no subdepartments (that is, the lowest level categories) on \href{www.amazon.com}{www.amazon.com} over a time period of several months in 2015-16.
\item{Hospital patient medical conditions:}
  We used a dataset of the number of patients per ICD-9 classified condition admitted to hospitals in the years 2000 to 2011.
  The source is the Observational Health Data Sciences and~Informatics (OHDSI) dataset~\cite{Hripcsak:2016fu}.
\item{German cities:}
  We used the population per each German city as of December 31st 2013, published by the Federal Statistical Office of Germany at
  \url{https://www.destatis.de/DE/ZahlenFakten/LaenderRegionen/Regionales/Gemeindeverzeichnis/Administrativ/Archiv/GVAuszugJ/31122013_Auszug_GV.xls?__blob=publicationFile}
\item{World population:}
  We used the world population for 2014 published by the World Bank, at the website \href{http://data.worldbank.org/indicator/SP.POP.TOTL}{http://data.worldbank.org/indicator/SP.POP.TOTL}
\item{US counties:}
  We used the population for each county of the United States, as of the 2010 census, as published on ``Annual Estimates of the Resident Population: April 1, 2010 to July 1, 2014, Source: U.S. Census Bureau, Population Division''.
\item{Languages:}
  We counted the number of native speakers for the languages of the world, published in the Encyclopedia of Language \& Linguistics, at the page \href{http://www.sciencedirect.com/science/article/pii/B0080448542090908}{http://www.sciencedirect.com/science/article/pii/B0080448542090908}
\end{enumerate}

\section{Representing the distribution}
To present the distribution of items among categories in these datasets
we use several different graphical representations.

One is the histogram of the number of categories with a given range of number of items (Figure~\ref{fig:histograms_sw}).
This can be interpreted as the probability mass function (PMF) of the data -- the probability that a randomly chosen category contains $n$ items.
This empirical PMF has two related shortcomings.
First, its shape depends on the choice of binning.
We use logarithmic binning, because the distribution is approximately lognormal for the well-sampled datasets.
Second, because of sparse sampling, the tails of the distribution are noisy even with logarithmic binning.

Another representation of the data is a \emph{rank-frequency} plot.
Namely, we rank the categories by the number of items that fall into each (rank 1 being the category with the most items), and plot the rank as a function of the number of items (Fig.~\ref{fig:allfreqrank}).
This graph is related to the cumulative distribution function (CDF) of the data: if a category has $n$ items, its rank is proportional to the number of categories that have a number of items greater or equal to $n$.
For this reason the rank-frequency plot is an unbiased estimator of the CDF, and does not rely on a binning.
Therefore, the tails of the distribution are accurately represented.

An alternative way to look at the data is to ask for the probability that a randomly chosen item will fall into a category with some given rank $r$.
The PMF of this probability is proportional to the quantile function, which is the functional inverse of the CDF -- in other words, the rank-frequency plot with horizontal and vertical axes interchanged.
This way of interpreting the data provides a robust estimator of a PMF, and we use it in comparing the model to the data.

\section{Random categorization model} \label{model}
As described in the main text, we introduce a random categorization model to explain the distribution of hierarchical taxonomies.
The model has no continuous or adjustable parameters, and the only inputs are the number $k$ of categories and the number $n$ of items present in the dataset.
It consists of the following steps:
\begin{enumerate}
\item
  Generate a random binary tree with $q \geq k$ leaves;
\item
  Assign a probability to each leaf;
\item
  Randomly populate the leaves with $n$ items according to these probabilities.
\end{enumerate}
For step 1, we begin from a single root category that we divide into two child categories.
Next, we choose one of these two categories at random and divide it into two, resulting in three categories.
At each successive time step, we choose a category at random (with uniform probability) and divide it in two.
We repeat this procedure until we reach the desired number of categories $q$ (``leaves'' of the
tree), which may be larger than $k$ (see below for details on how $q$ is determined).

To each leaf, we assign a probability, where $b$ is the depth of the leaf (the number of divisions separating that leaf from the root), $x$ is a uniform random variable between ­$-1$ and $1$, and $x$ is a normalization factor that ensures the probabilities sum to one.
The factor corresponds to an equal expected division of the items in a parent category into each child category.
Adding the random variable $x$ prevents the frequencies of items in well-sampled datasets from being artificially peaked at integer powers of two.
We could also achieve a similar ``smeared'' distribution by assigning probabilities to each child category according to some continuous distribution rather than equally, or by dividing categories into a randomly varying number of child categories.

Of course, $b$ is itself a random variable that depends on the particular leaf and the structure of the random tree.
Given a random tree with $q$ leaves, the probability that a randomly chosen leaf has depth $b$ is given by \cite{Drmota:2009tm}
\be
    P(b, q) = 2^{-b} \frac{C(q-1, b)}{\Gamma(q+1)} \, .
    \label{eq:P}
\ee
Here \emph{C} is an unsigned Stirling number of the first kind, which obeys the recursion relation
\be
    C(n+1, k) = n C(n, k) + C(n, k-1)
\ee
with boundary conditions $C(n,0)=0$ for $n \geq 0$ and $C(n,k)=0$ for $k \geq 0$, with $C(n,n)=1$ for $n \geq 0$.
From this and the definition of the Gamma function, we derive a recursion relation for the probability:
\be
    P(b, q+1) = \frac{q-1}{q+1} P(b,q) + \frac{2}{q+1} P(b-1, q) \, ,
\ee
with boundary conditions $P(0,1)=1$, $P(0,q)=0$ for $q \geq 0$.
This recursion relation allows an efficient numerical calculation of the probabilities for arbitrary $b$, $q$, values.

This distribution is approximately normal in $b$ near its maximum, with variance $\sig_q^2 = \sum_{j=2}^q (2/j - 4/j^2) \simeq -3.4 + 2 \ln q$ and mean $\mu_q = \sum_{j=2}^q 2/j \simeq -0.85 + 2 \ln q$, where the approximations are valid at large $q$.
Since the probabilities $p$ are related to $b$ by $p=2^{-b}$, the variance in $\ln p$ is
\be
    \sig_{\ln p}^2 \simeq (\ln 2)^2 (-3.4 + 2 \ln q) \, .
    \label{eq:sigma2_lnp}
\ee

Note that if $q$ is the number of species, the functional relation between the variance and the number of species is similar to the canonical lognormal~\eqref{eq:canln} in that the logarithm of the number of species is linear in the variance at leading order, albeit with a different coefficient.
This is a general feature of random tree models~\cite{Sugihara:1980bx}.

Having generated a tree and assigned probabilities, the last step is to assign $n$ items to the leaves by distributing them according to a multinomial distribution of $n$ over the $q$ probabilities $p_i$.
There remains the question of determining the number of leaves $q$ that the tree should have.
To do so, we notice that, given a tree with $q$ leaves, after multinomial sampling with $n$ items, we expect to get the following fraction of empty categories:
\be
    f_e(q, n) = \sum_{b=1}^q P(b,q) (1-2^{-b})^n \, ,
\ee
with $P(b,q)$ given by eq.~\ref{eq:P}.
If we wish to obtain an expected number $k$ of nonempty categories, we should choose $q$ by solving the nonlinear equation $k = q (1 - f_e(q,n))$.
We use the value $q(k,n)$ that solves this equation for the number of categories we generate to model a dataset with given $k$, $n$.

These steps can be thought of as generating a random partition of $n$ into $q$.
The probability distribution of these random partitions does not have a simple analytical expression as far as we are aware, but it is not difficult to sample numerically.
To do so, we generate many random trees following the previous steps.
For the datasets in which $q \gg k$, we only keep those trees where after populating with $n$ items, the number of nonempty categories within $1 \%$ of $k$ (ideally we could keep only those with exactly the same number of non-empty categories as the data, but this would be too time-consuming and in any case does not affect the results significantly).
For each dataset, we generate 100 such random partitions.

The code is implemented in Python, and it is available upon request from the authors.
We have also made a web application available at the address \href{https://rabadan.c2b2.columbia.edu/universaltaxdist}{https://rabadan.c2b2.columbia.edu/universaltaxdist} in which the user can test the predictions of our model against any dataset of counts.

\section{More general random subdivision models}
There are many natural generalizations of this model.
One is to assign unequal probabilities to the subcategories at each division.
That is, for the binary model described above we assigned probabilities $p = 2^{-b+x}/X$ to the
leaves, where $b$ is the number of branchings separating that leaf from the root and $x$ is a random variable ranging from $-1$ to $1$.
Apart from the smearing factor $x$, this corresponds to an equal (50-50) split at each binary subdivision.
Instead, consider assigning probabilities to each of the two subcategories produced at a division according to some probability distribution $\rho(x)$ with $0 \leq x \leq 1$ (for instance, $\rho(x)$ could be a beta distribution), rather than necessarily equally.
In this case, the variance of the log probability for a random leaf is the variance in the number of divisions $b$ times a factor of $\avg{(\ln x)^2} = \int_0^1 (\ln x)^2 \rho(x) \rmd x$.
Note that for the 50-50 model, $\avg{(\ln x)^2} = \( \ln \frac{1}{2} \)^2 = \(\ln 2 \)^2$ (see~\eqref{eq:sigma2_lnp}).
One can easily show that $(\ln 2)^2$is the minimum possible value of $\avg{(\ln x)^2}$ for distributions satisfying $\int_0^1 \rho(x) \rmd x = 1$ and $\rho(x) = \rho(1-x)$, as is necessary here.
This shows that the model we have considered has the minimum possible variance within a large class of random division models.

Another generalization is to divide into $j$ subcategories according to some set of probabilities $p_j$, producing trees that are more general than the simple binary ($j=2$ only) structure considered so far.
For a model that always divides into a fixed number $j$ of subcategories (producing a ``$j$-nary'' tree rather than a binary tree), the mean and variance of $b$ (the number of branchings separating a random leaf from the root) for a tree with q leaves are
\be
    \mu_j(q) = \frac{j}{j-1} \[ H\(\frac{q}{j-1}\) - H\(\frac{1}{j-1}\) \] \, ,
\ee
where $H$ is a harmonic number, and
\begin{multline}
        \sig_j^2(q) = \frac{j}{(j-1)^2} \bigg[ (j-1) \( \psi^{(0)}\( 1+ \frac{q}{j-1}\) - \psi^{(0)}\(2+\frac{1}{j-1}\) \) + \\
    j \( \psi^{(1)}\( 1+ \frac{q}{j-1}\) - \psi^{(1)}\(2+\frac{1}{j-1}\) \) \bigg] \, ,
\end{multline}
where $\psi^{(m)}(z)$ is a polygamma function of order $m$. If the probabilities are equally divided among the child categories so that $p = j^{-b}$, the mean and variance of the log probabilities are related to the above results simply by factors of $\ln j$ and $(\ln j)^2$~\cite{Bowler:2012ce} respectively (see~\eqref{eq:sigma2_lnp}).
For $q \gg j$, the result is a variance in the log probabilities that is an increasing function of $j$.
Hence, one sees that the minimum variance for a tree with a given number of leaves is achieved for the case of binary trees ($j=2$).

A third generalization would be to choose to divide categories with a probabilities depending on the number of items in them (or the probability so far assigned to them), rather than uniformly.
That is, one might consider a model where at each step large categories are more likely to be subdivided than small~\cite{Agrawal:2007if,Drmota:2009tm}.
This can produce an item distribution with lower variance than the model presented here, but we will not consider it further here.

We have chosen to focus on the binary ``equal divide'' model because we regard it as the simplest, and it seems to do a surprisingly good job fitting the data.
We leave to future work the very interesting question of whether more complex or alternative models might work even better.

\section{Comparison between theory and data} \label{comp}
We should think of the partition of items provided by the data as a single random sample of some underlying distribution.
It is this underlying distribution that our model attempts to reproduce.
In this situation it is difficult to use a null hypothesis frequentist test.
A Bayesian inference approach would strongly depend on the chosen prior.
It is possible to do a non-parametric test based on the empirical CDF, as for instance the Kolmogorov-Smirnov, Anderson-Darling or Mann-Whitney test.
These are however sensitive to particular aspects of the frequency distribution, and would not properly express the degree by which our statistical model can capture the observed partition of items.

For these reasons, we characterize the degree of agreement of the data with our model (or the best-fit lognormal) by calculating the Kullback-Leibler (KL) divergence, or relative entropy, between our model and the data.
In general, given two probability distributions $p$ and $q$, the KL divergence is defined by
\be
    D_{KL}(p || q) = \int_{-\infty}^{+\infty} \rmd x \, p(x) \log \frac{p(x)}{q(x)}
\ee
with a sum replacing the integral in the discrete case.
This quantity is not symmetric in $p$ and $q$.
Typically, $p$ represents a ``fiducial'' distribution, while $q$ is a model for or approximation of $p$.
In the context of information theory, $D_{KL}(p||q)$ is the amount of information lost when the distribution $q$ is used to approximate $p$.

This approach to model comparison was used recently in ecology in the context of species-abundance distributions~\cite{Alroy:2015hh}.
There, $x$ is the rank of the species and $p$ and $q$ are estimates of the probability that a given individual belongs to the species with rank $x$ in the data and model respectively.
Here, we employ a methodology closely related to that used by Alroy~\cite{Alroy:2015hh}.
To calculate we estimate the probability that a random item belongs to a category with rank $x$ for both data and model.
We estimate the probability for the data $p(x)$ and for each sampled tree $q(x)$ simply by the normalized fraction of items in the category with rank $x$.
That is, $p(x)=n(x)/n$, where $n(x)$ is the number of items in the category of rank $x$ and $n$ is the total number of items, and the same for $q(x)$.
We then compute $D_{KL}(p||q)$ using these estimates.

To compare our model to the data, we calculate the KL divergence for a set of 100 sampled trees, and then plot a histogram (actually, a kernel density estimate) of the results, showing the median KL divergence with a dashed vertical line.
As a comparison, for each dataset we also consider the lognormal distribution with mean and variance fit to the data $p(x)$ using maximum likelihood estimators.
Note that this lognormal distribution has two continuous parameters, in addition to the discrete values $k$ and $n$ required to model the data.
For each dataset we generate 100 sets of $k$ draws $q(x)$ from the lognormal distribution.
Each set of draws is then normalized so that $q(x)$ sums to 1, and we compute the KL divergence of each of these 100 sets relative to the data.
The results are plotted in Fig.~\ref{fig:allfreqrank}.

As measured by the KL statistic, for some datasets the lognormal fit is better, while for others the model is superior (Fig.~\ref{fig:allfreqrank}).
However, as mentioned above, lognormal has two continuous parameters that are fit to the data, while the model has none.
As such it is remarkable how well it fits this wide variety of classification systems.

\end{appendices}

\begin{table}[h]
\begin{adjustbox}{center}
\begin{tabular}{|m{0.3\textwidth}||r|r|r|r|r|r|}
    \hline
Dataset & $k$ & $n$ & $n/k$ & $q/k$ & ${\rm Chao}/k$
& $W$ \\
    \hline
Books per LOC call number in Harvard University library's collection &
1,053,886 & 4,870,797 & 4.622 & 37.98 & 2.602 & 0.647\\

Files per folder on a laptop & 228,972 & 1,347,541 & 5.885 & 15.96 &
1.290 & 0.785\\

Species abundance in human oral microbiome & 1,179 & 34,753 & 29.48 &
1.444 & 1.572 & 0.866\\

Books per NATC number published nationwide & 631 & 13,672,817 & 21669 &
1.000 & 1.000 & 0.932\\

Proteins per CATH category & 2,738 & 235,858 & 86.14 & 1.264 & 1.071 &
0.964\\

Articles per subject on the ArXiv & 144 & 61,585 & 427.7 & 1.008 & 1.000
& 0.974\\

Lost and found objects in the New York subway & 216 & 539,183 & 2496 &
1.002 & 1.000 & 0.991\\

Items per department on Amazon.com & 51,320 & 397,797,849 & 7751 & 1.039
& 1.007 & 0.996\\

Number of patients per medical condition & 10,695 & 606,097,666 & 56671
& 1.003 & 1.008 & 0.996\\

Population of German cities & 11,161 & 80,767,463 & 7237 & 1.019 & 1.000
& 0.996\\

Population of US counties & 3142 & 308,758,105 & 98268 & 1.001 & 1.000 &
0.992\\

World population & 214 & 7,184,343,030 & 33571696 & 1.001 & 1.000 &
0.967\\

Languages of the world & 6458 & 5,719,611,222 & 885663 & 1.000 & 1.005 &
0.995 \\
\hline
\end{tabular}
\end{adjustbox}
\caption{The datasets used in this analysis. The number of categories in each dataset is $k$, the total number of items is $n$, and the number of categories in the tree used to model the data is $q$, so that \emph{q/k} is our model's estimate of the ratio of the total number of categories to the number of non-empty categories.
We compare this to the Chao estimator (``\emph{Chao/k}'') of the same ratio, and also give the Shapiro-Wilk \emph{W} normality statistic for the logarithm of the counts distribution.}
\label{tab:data}
\end{table}

\begin{figure}
    \centering
    \begin{subfigure}[t]{0.9\textwidth}
        \includegraphics[width=\textwidth]{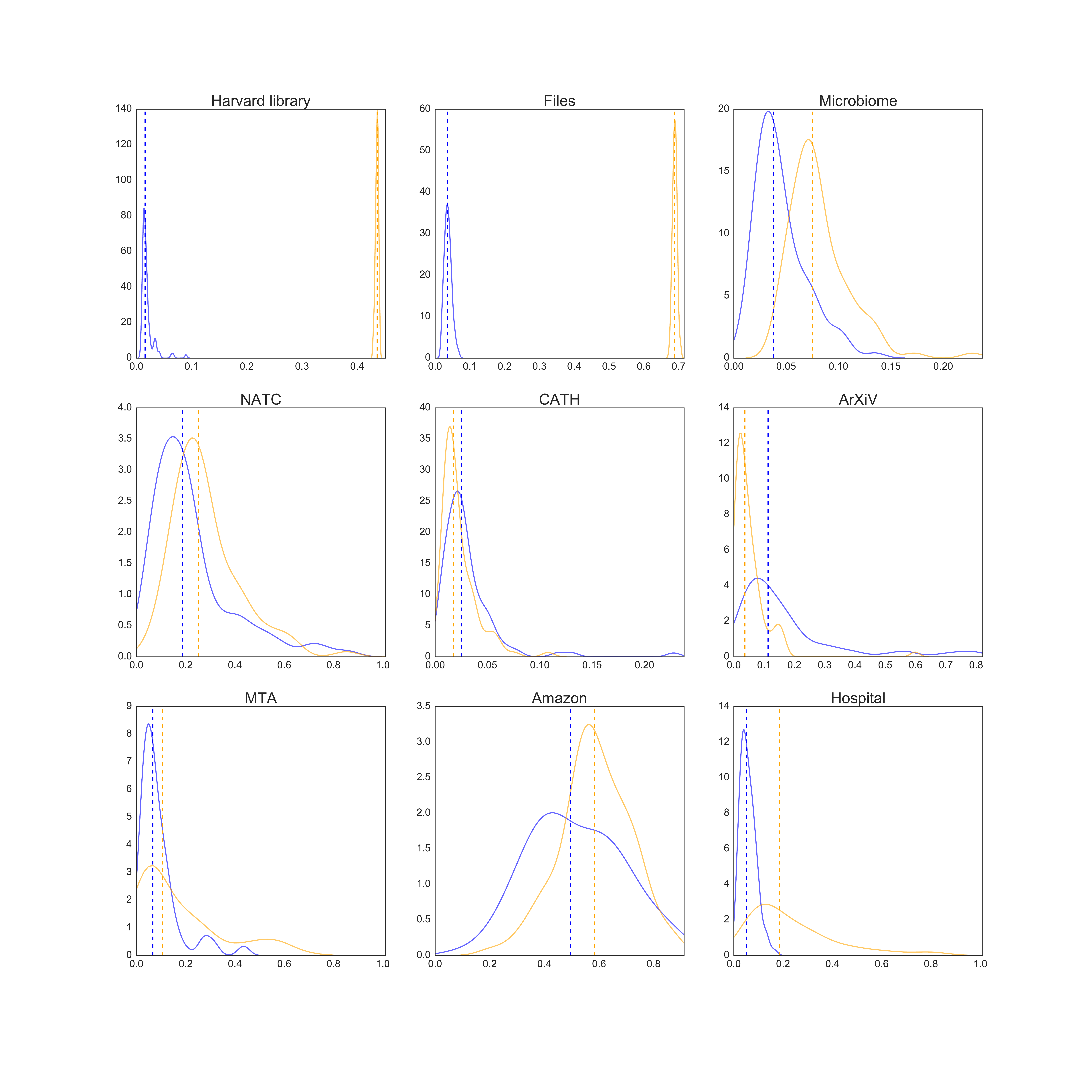}
        \label{fig:KL}
    \end{subfigure}
    \phantomcaption  
\end{figure}
\begin{figure}\ContinuedFloat
    \centering
    \begin{subfigure}[t]{0.9\textwidth}
        \includegraphics[width=\textwidth]{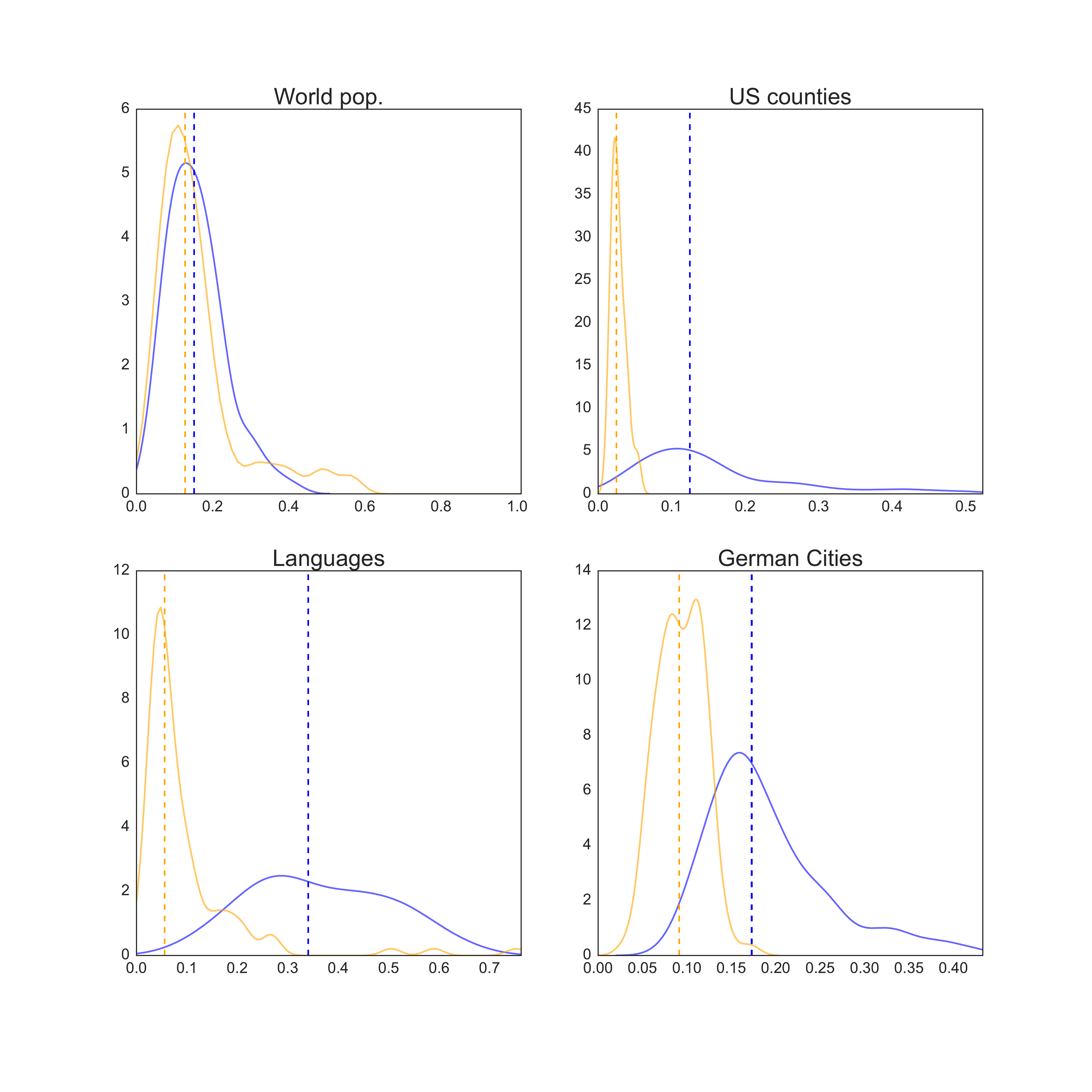}
        \label{fig:demoKL}
    \end{subfigure}
    \caption{Model comparison using the Kullback-Leibler statistic.
    Plotted are kernel density estimates of histograms of the estimated Kullback-Leibler divergence with respect to the data, from 100 realizations of the random model (blue curves) or 100 sets of $k$ samples from the best fit lognormal distribution (orange curves).
    The median KL value is marked with a dashed vertical line.
    Smaller KL values correspond to a better fit.
    Dashed vertical lines indicate the median value.
    The plots in the top panel are the hierarchical datasets (first nine entries, Table~\ref{tab:data}); the plots in the bottom panel are the four non-hierarchical datasets (last four entries, Table~\ref{tab:data}).}
    \label{fig:allKL}
\end{figure}

\begin{figure}
    \centering
    \begin{subfigure}[t]{0.9\textwidth}
        \includegraphics[width=\textwidth]{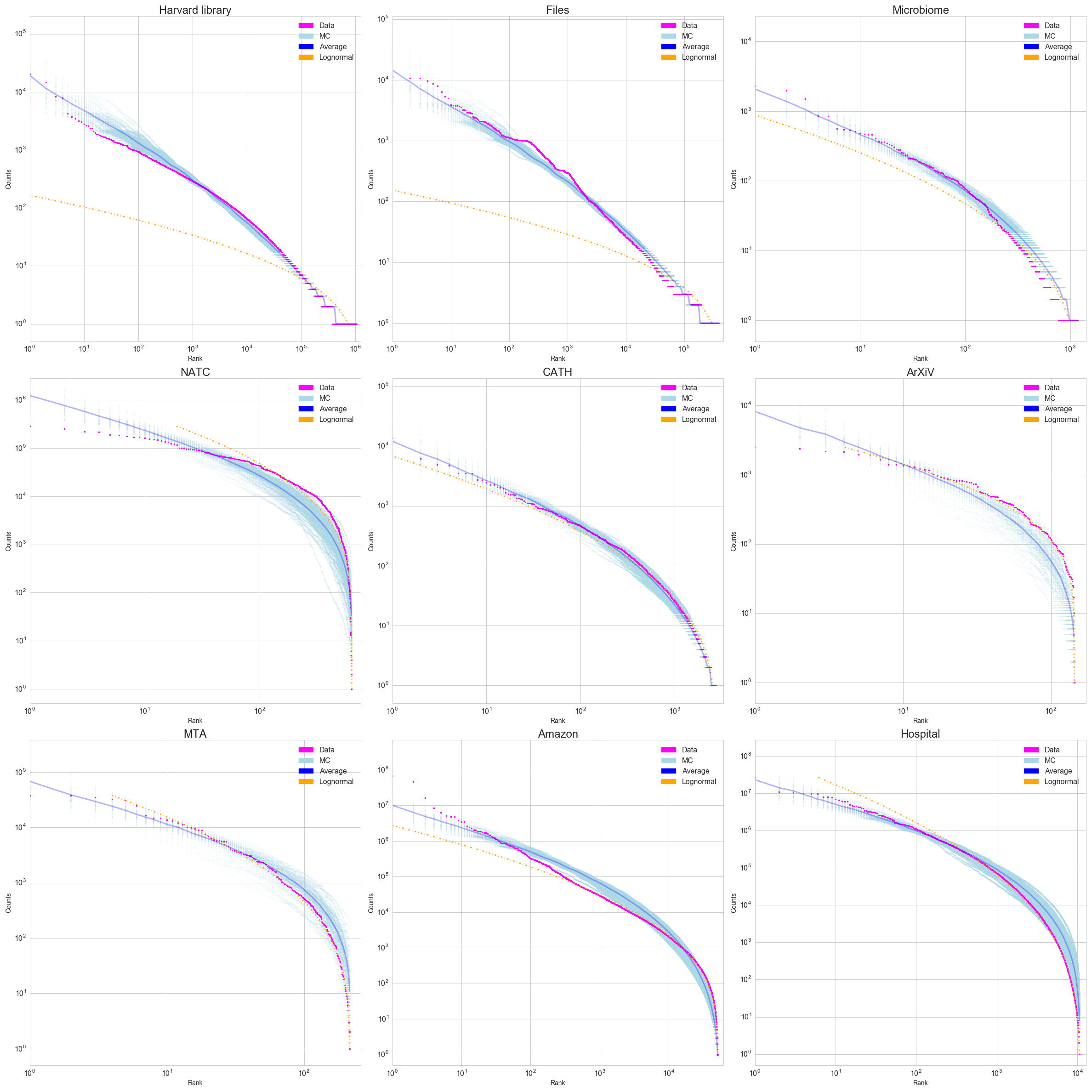}
        \label{fig:freqrank}
    \end{subfigure}
    \phantomcaption  
\end{figure}
\begin{figure}\ContinuedFloat
    \centering
    \begin{subfigure}[t]{0.9\textwidth}
        \includegraphics[width=\textwidth]{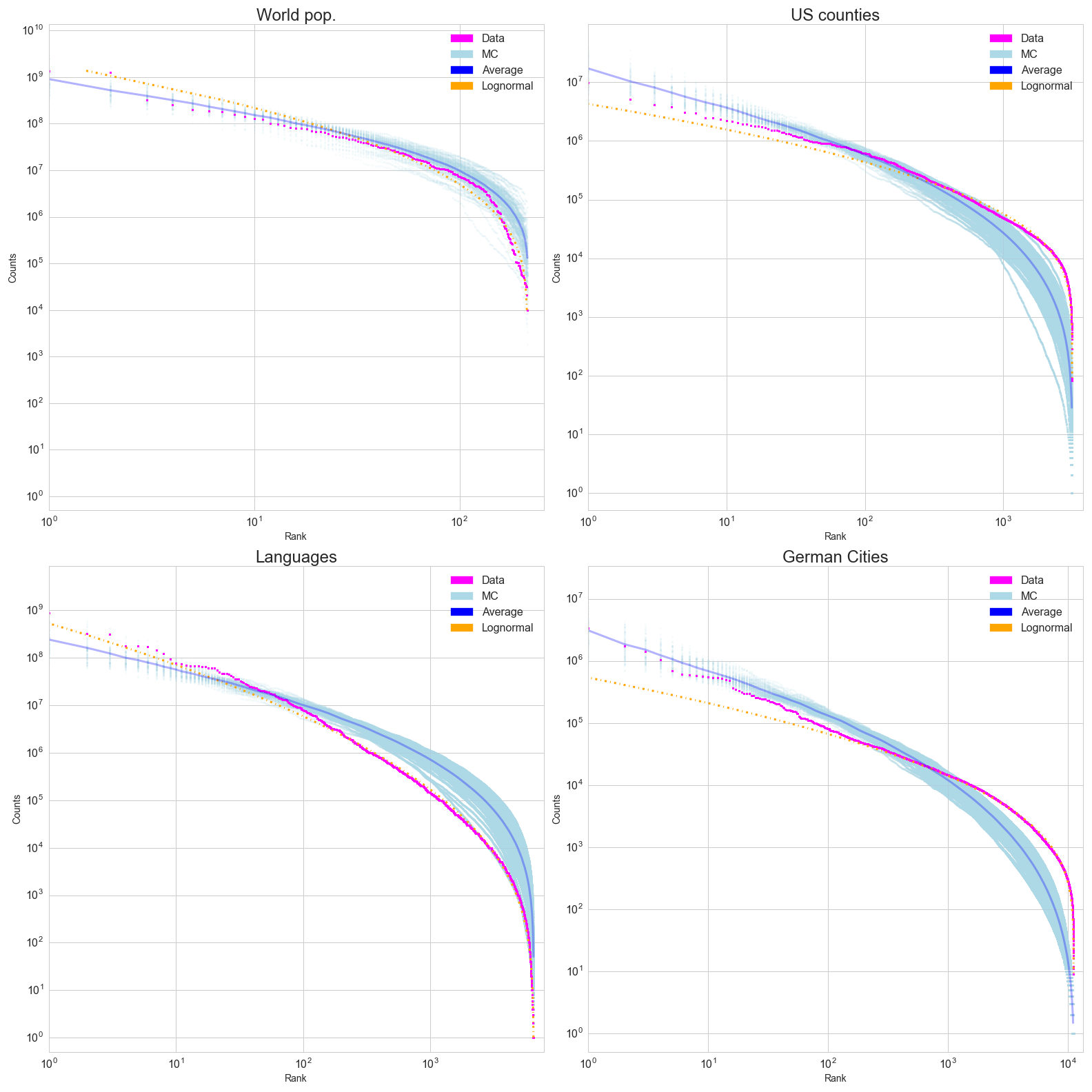}
        \label{fig:demofreqrank}
    \end{subfigure}
    \caption{Rank-abundance plots for each dataset.
    Magenta points are data, light blue points are 100 realizations of the random model, whose average is the blue curve.
    Orange is the best fit lognormal.
    The plots in the top panel are the hierarchical datasets (first nine entries, Table~\ref{tab:data}); the plots in the bottom panel are the four non-hierarchical datasets (last four entries, Table~\ref{tab:data}).}
    \label{fig:allfreqrank}
\end{figure}

\bibliography{refs}

\end{document}